\documentclass[%
 aip,
 reprint,%
]{revtex4-1}

\usepackage{graphicx}
\usepackage{dcolumn}
\usepackage{bm}

\usepackage[utf8]{inputenc}
\usepackage[T1]{fontenc}
\usepackage{mathptmx}
\usepackage{etoolbox}
\usepackage{color}

\usepackage[normalem]{ulem}
\usepackage{mathrsfs}
\usepackage{placeins} 
\usepackage{graphics}
\usepackage{bm}
\usepackage{subfigure}
\usepackage{graphicx}
\usepackage{amsthm}
\usepackage{notes2bib}
\usepackage{amsmath}
\usepackage{amssymb}
\usepackage{url}
\usepackage{enumerate}
\usepackage{color}
\usepackage{mathtools}
\usepackage{setspace}
\usepackage{color}

\begin{document}
\title{Time-delayed feedback control for random dynamical systems}

\author{Miki U. Kobayashi}
\affiliation{Faculty of Economics, Rissho University, Tokyo 141-8602, Japan}
\author{Yuzuru Sato}
\affiliation{Department of Mathematics, Hokkaido University,  Kita 12 Nishi 7, Kita-ku, Sapporo, Hokkaido 060-0812, Japan}
\affiliation{London Mathematical Laboratory, 8 Margravine Gardens, Hammersmith, London W6 8RH, UK}

\begin{abstract}

We extend the Pyragas time-delayed feedback control (TDFC) to apply it to random dynamical systems and introduce an extended  classification based on Lyapunov exponents and trajectory fluctuations. We demonstrate the applicability of this framework using the random logistic map and the stochastic Rössler system. Our results reveal that noise-induced chaos triggers a transition from stable to unstable regimes based on a phenomenon inherent to random dynamical systems.
\end{abstract}
\date{\today}
\maketitle
\begin{quotation}
Chaos control aims to stabilize deterministic chaotic dynamical systems, effectively converting chaotic motion into periodic behavior. Among the most widely studied techniques is the time-delayed feedback control (TDFC) method introduced by Pyragas \cite{Pyragas1992}, which stabilizes unstable periodic orbits (UPOs) embedded in a given deterministic strange attractor.
In this work, we extend the TDFC framework to random dynamical systems by introducing an extended classification that quantifies whether a given control scheme successfully stabilizes the system under stochastic perturbations. We demonstrate the effectiveness of our approach through applications to random logistic maps and stochastic Rössler systems. 
Our results show that the transition from controllable to uncontrollable regimes in time-delayed feedback control is driven by noise-induced chaotic behavior, in which increasing noise levels overwhelm the deterministically stable periodic orbit and generate a random strange attractor with a positive Lyapunov exponent.
\end{quotation}
\section{Introduction}

The aim of chaos control is to stabilize deterministic chaotic dynamical systems, 
effectively transforming chaotic motion into periodic behavior \cite{Romeiras1992,Shinbrot1993}.
The time-delayed feedback control (TDFC), introduced in 1992 by Pyragas, is one of the most notable methods for chaos control by stabilizing unstable periodic orbits (UPOs) embedded in chaotic attractors \cite{Pyragas1992}. 
Many researches have been conducted theoretical \cite{Bielawski1994,Just1997, Just1998} and applied \cite{Postlethwaite2009,Kobayashi2012} research on TDFC, as well as  significant practical contributions \cite{Nakajima1997,FiedlerPRL,Yamasue2006} 
have been made.
The key idea is very simple yet powerful: inject a corrective signal derived from the difference between the current system state and its state periods earlier. Since the feedback vanishes upon stabilization of the target periodic orbit, the method is both non-invasive and reference-free. 
The time-delayed feedback control for deterministic chaotic systems is a kind of controlling chaos study, and 
its purpose is to transform chaotic motion into periodic behavior. A Criteria for the success of the control by time-delayed feedback control for deterministic chaotic systems 
is that when the largest Lyapunov exponent is negative, the control is successful; when it is positive, the control fails. 

In many natural and engineered systems, uncertainty and stochastic perturbations play a crucial role in determining long-term behaviour. Traditional deterministic dynamical systems often fail to capture such phenomena fully, which has led to the development of Random Dynamical Systems (RDS) as a natural extension of classical theory. RDS combines tools from ergodic theory, stochastic analysis. Random maps and  stochastic differential equations in order to study the evolution of systems under the influence of noise. Arnold established a formal framework for RDS, providing a rigorous setting in which to analyze stability, attractors and bifurcations in stochastic environments \cite{Arnold1998}.
A key concept in RDS is random attractor: a random compact set $A(\omega)$ with $\omega$ which can be interpreted as a noise realization in practice. It is invariant under the cocycle and attracts all bounded sets almost surely (See Appendix for the details). 
Random attractors generalize the notion of global attractors from deterministic systems, and are essential for describing long-term behavior of dynamical systems in the presence of noise. 
Furthermore, stochastic bifurcation theory has emerged as a powerful tool for studying qualitative changes in dynamics induced by variations in system parameters or noise intensity.

In this paper, we focus on RDS with the time delayed feedback, i.e. delayed random dynamical systems. 
It is found that there are many interesting phenomena in delayed random dynamical systems, e.g. stochastic resonance caused by a interaction between time delay and noise \cite{Ohira1999}. 
To the best of the author's knowledge, there are few examples of TDFC applied to random chaotic dynamical systems \cite{Janson2004,Balanov2004}, 
and it is not clear whether random chaotic dynamical systems can be controlled using TDFC. 
If it is possible, how the criteria should be defined.
In the case of time-delayed feedback control for random dynamical systems, not only Lyapunov exponents but also variance of a controlled trajectory are required as the  criteria. 
In this paper, we define a stabilization classification for random dynamical systems with time-delayed feedback based on a Lyapunov exponent
$\lambda$ 
(see Appendix A), the ratio of the standard deviation of the trajectory $\sigma$, and of the external noise $\epsilon$. The three regimes are naturally introduced as follows:\\

\newpage

\noindent
i) Strongly stable (SS)

\[ { \sigma \over \epsilon }\propto  O(1), ~~\lambda <0
\]
The Lyapunov exponent is negative, 
and the fluctuations of the controlled orbit remain comparable to those of the external noise.\\

\noindent
ii) Weakly stable (WS)

\[ { \sigma \over \epsilon } \gg  O(1), ~~\lambda < 0
\]
Although the Lyapunov exponent is negative, 
the trajectory exhibits larger variability than the system noise.
In this case, we typically observe switching dynamics between the target periodic orbit and unstable chaotic set. 
We quantify the sojourn-time rate $p$
near the periodic orbit in the observed dynamics. 
Based on this rate, we define weakly stable dynamics as 
$p$-stable.
Strong stability corresponds to 
$p\simeq 1$, while instability yields 
$p\simeq 0$.
\\

\noindent
iii) Unstable (U)

\[ { \sigma \over \epsilon } \gg  O(1),  ~~\lambda > 0\]
A positive Lyapunov exponent indicates the emergence of a random strange attractor and the failure of control.


Note that in the case of time continuous random dynamical systems, 
a zero Lyapunov exponent is not guaranteed to exist, 
and in fact, it is typically absent. Therefore, throughout this paper, 
when considering the strongly stable control (SS) and the weakly stable one (WS), 
we use the second Lyapunov exponent, i.e. the maximal Lyapunov exponent excluding the first exponent, that is the null exponent in deterministic flows. 
In the case of the strongly stable controlled dynamics (SS), the largest Lyapunov exponent (the second exponent in the case of time continuous systems) is negative, and fluctuation of trajectory can be suppressed to be equivalent to one of noise by the feedback control.
In the case of the weakly stable (WS), the largest Lyapunov exponent (the second exponent in the case of time continuous systems) is negative, but the standard deviation of the trajectory is larger than that of the noise.
The unstable (U) is defined by the positive largest Lyapunov exponent.

From the viewpoint of random dynamical systems, the attractor observed in the SS regime corresponds to a pseudo-periodic orbit, while that in the WS regime corresponds to a partially chaotic orbit \cite{Thai2018, Sato2018}. In contrast, the U regime is characterized by a random strange attractor with a positive Lyapunov exponent (see the Appendix for details). 

The transition points from SS to  WS can be characterized by the dichotomy spectrum \cite{Sato2018} in  special cases, which is interpreted as the support of the distribution of possible finite time Lyapunov exponents.  However, it is governed by rare noise realizations and therefore cannot be sharply identified experimentally.   
We give the definitions by measurable quantities, and they expect to be applicable to a broad class of systems, including experimental settings.

In Section II, we apply the TDFC method to the stochastic logistic map 
and show the advantages of our criteria for random dynamical systems. 
In Section III, we analyze the stochastic R\"ossler system cases.
Finally, in Section IV, a summary and discussion are presented.

\section{Time-delayed feedback control for the random logistic map}
We consider the random logistic map with time-delayed feedback:
\begin{eqnarray}
{x}_{n+1}=a-{x_n}^2+D\xi_x+K(x_{n-\tau}-x_n),
\end{eqnarray}
where $\xi$ is uniform noise $-1 < \xi < 1$.
It is known that in the case of the logistic map without noise, the time-delayed feedback makes a chaotic trajectory  periodic. 
In this paper, we clarify whether the time-delayed feedback control can be applied to the logistic map with noise.

Eq. (1) can be transformed into the following generalized Henon map:
\begin{eqnarray}
X_{n+1}=A-X_n^2+X^{(\tau)}_n+D \xi_n,\\
X^{(i)}_{n+1}=X^{(i-1)}_n, i=1,2,3,\cdots \tau-1, \\
X^{(\tau)}_{n+1}=KX^{(\tau-1)}_n,
\end{eqnarray}
where $\displaystyle{X_n=x_n+{K \over 2}, X^{(0)}_n=x_n,A=a+{K(K+2)\over 4}}$.
We numerically calculate Eq. (2) $\sim$ (4) for random logistic map with time-delayed feedback. 
In this paper, 
we fix the parameter values $a=1.784, \tau=3, K=0.05$ and change the noise intensity $D$. 
Note that the standard deviation of system noise $D \xi_n$ is $\displaystyle{{2D \over \sqrt{12}}}$.

\begin{figure}[ht]
    \begin{tabular}{cc}
    \includegraphics[scale=.35]{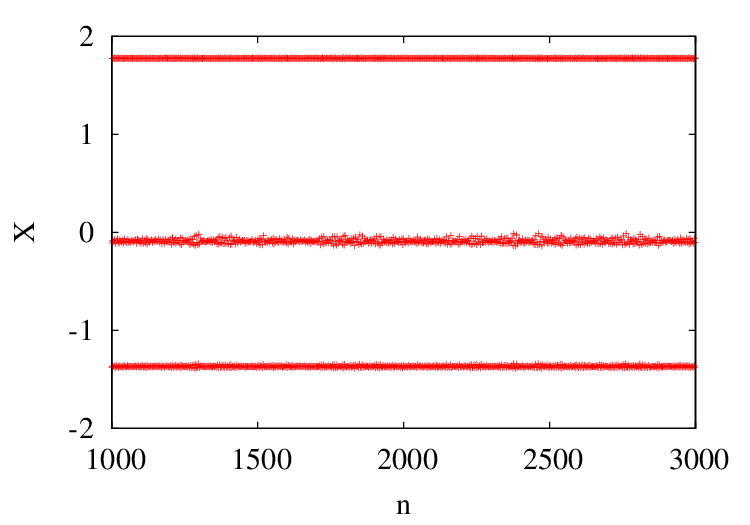} &
    \includegraphics[scale=.35]{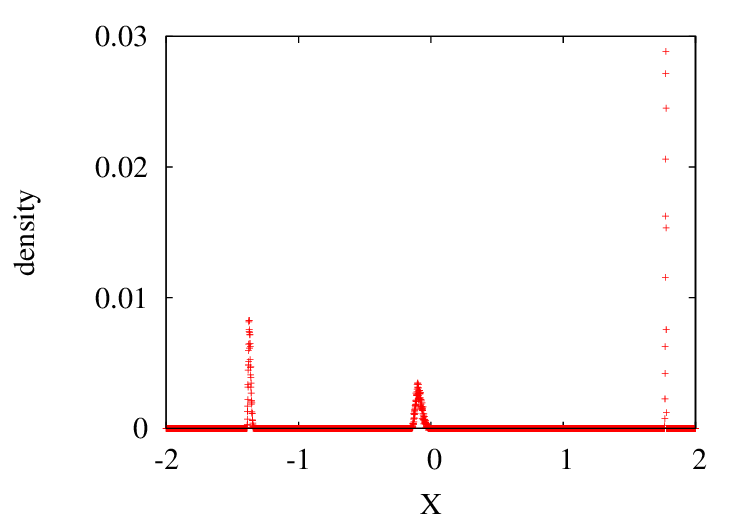} \\     
      \end{tabular}

    \caption{\label{kobayashi_fig:fig1}
    $a=1.784, D=0.002, K=0.05, \tau=3$
      (left) Stabilized pseudo-period-3 orbit.
      (right) 
      The density distribution function for time series $x$.
      The supports of the distribution function is disconnected.  
      The standard deviation of trajectory is $0.0024$, which is approximately the same order of the standard deviation of noise $0.00115$. The largest Lyapunov exponent is $-0.02$.
      This case corresponds to the strongly stable control.
      }       
\end{figure}

\begin{figure}[ht]
    \begin{tabular}{cc}
          \includegraphics[scale=.35]{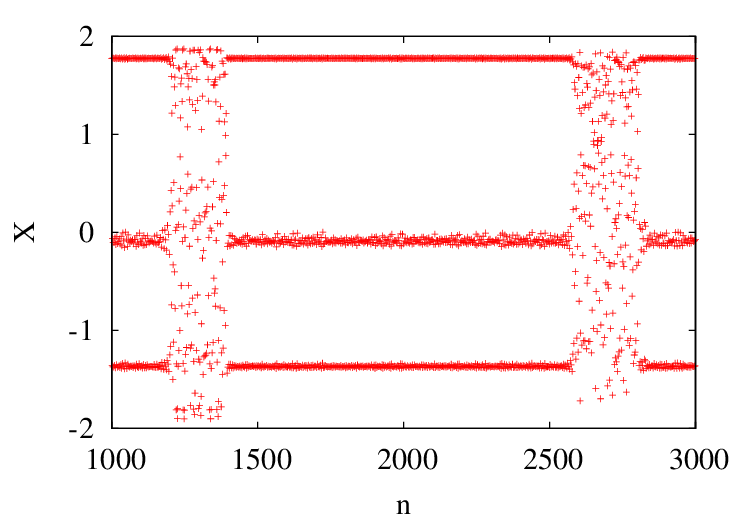}  &
      \includegraphics[scale=.35]{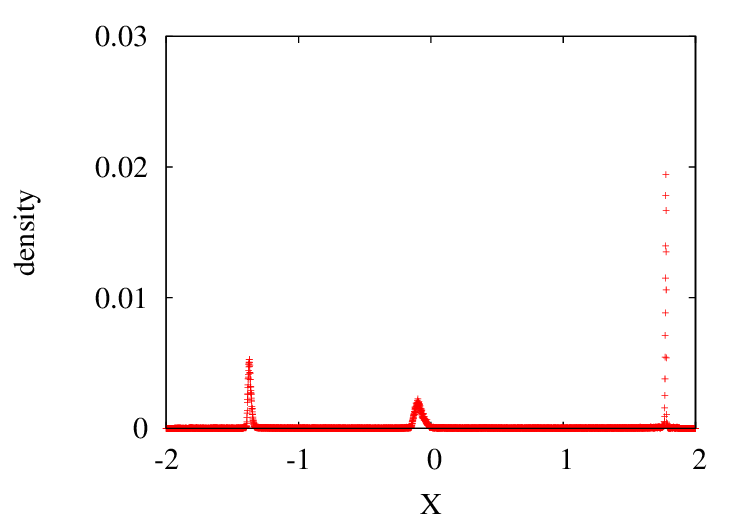} 
      \end{tabular}

    \caption{\label{kobayashi_fig:fig1}
     $a=1.784, D=0.00265, K=0.05, \tau=3$
      (left) Partially chaotic orbit.
      (right) 
      The density distribution function for time series $x$.
      The supports of the distribution function is connected.
      The standard deviation of trajectory is $1.198$, which is larger than the standard deviation of noise 0.00153, still the largest Lyapunov exponent is $-0.007$.
      This case corresponds to the weakly stable control (WC) with $p=0.66$. The value $p$ represents the proportion of time during which $|x_n - x_{n - \tau}|$ does not exceed the standard deviation of the system noise.
      }
\end{figure}

\begin{figure}[ht]
    \begin{tabular}{cc}
          \includegraphics[scale=.35]{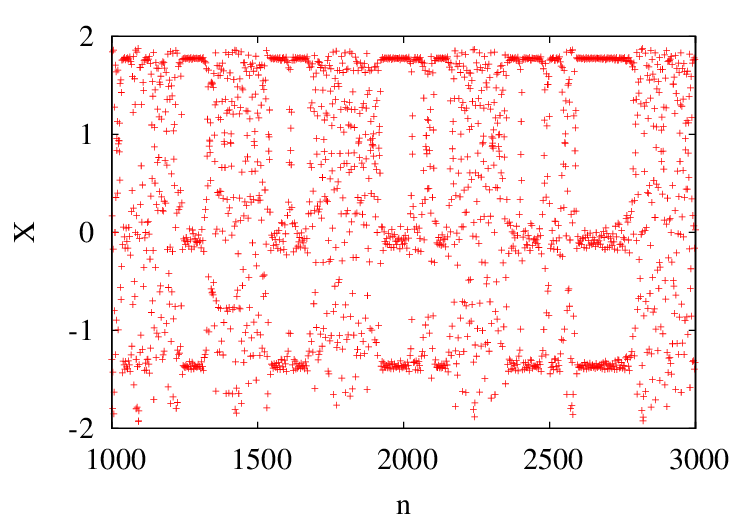}  &
      \includegraphics[scale=.35]{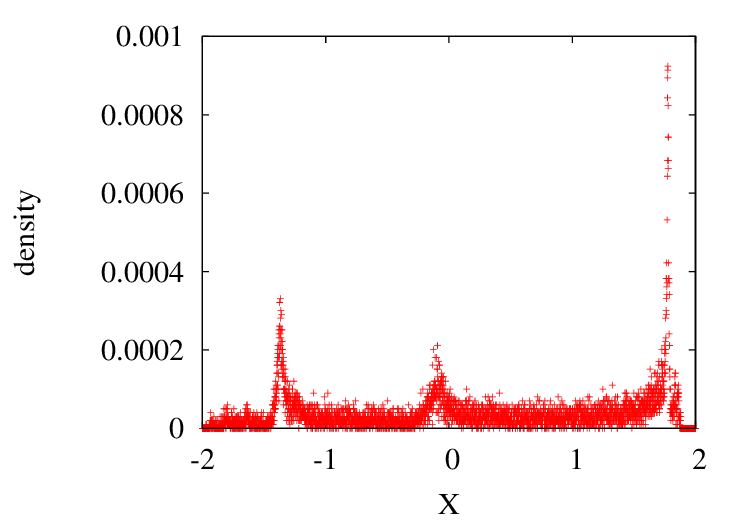} 
      \end{tabular}

    \caption{\label{kobayashi_fig:fig1}
     $a=1.784, D=0.003, K=0.05, \tau=3$
      (left) Random strange attractor.
      (right) 
      The density distribution function for time series $x$.
      The supports of the distribution function is connected.
      The standard deviation of trajectory is $1.177$, which is larger than the standard deviation of noise 0.0025, the largest Lyapunov exponent is  $0.014$.
      This case corresponds to the unstable control.
       }       
\end{figure}

We control chaos near the period-3 window to periodic points with period 3 by using control with delay time $\tau=3$.
Fig. 1 shows the time series of the logistic map (left) and its stationary distribution function(right) using a=1.784, D=0.002, K=0.05, tau=3. The standard deviation of the trajectory is 0.0024, which is almost on the same order of magnitude as that of the noise.
In addition, the Lyapunov exponent is negative at -0.02, which means that the classification in our new definition is the strongly stable control (SS).

Fig. 2 shows the time series (left) and its stationary  distribution (right) of the logistic map with D=0.00265.
The standard deviation of the trajectory is 1.198, which is large compared to the standard deviation of the noise.
The Lyapunov exponent is negative at -0.007, which means that the classification is weakly stable (WS) under our definition.
The orbit appears to be intermittently chaotic and switching dynamics between the target
periodic orbit and unstable chaotic set is observed. 
Since the Lyapunov exponent is negative, it is a partially chaotic orbit \cite{Thai2018, Sato2018}.
Note that the finite time Lyapunov exponent in the burst regions of this orbit can be positive. 


Fig. 3 shows the time series of the logistic map (left) and its stationary  distribution function (right) using D=0.003.
The standard deviation of the trajectory is 1.177, which is large compared to that of the noise. The Lyapunov exponent is positive at 0.014, making it unstable (U) by our new definition of classification.

In the case of $a=1.784, K=0.05$, the controlled system with noise intensity $0 < D < 0.0023$ is stable, one with noise intensity $0.0023 \le D < 0.0027$ is weakly stable, and one with noise intensity $0.0027 \le D$ is unstable.  
In fact, Fig. 4 shows the change in the standard deviation and the largest Lyapunov exponent for different values of D. It can be seen that there is a transition from strongly stable (SS) to weakly stable (WS) around D=0.0023 and from weakly stable (WS) to unstable (U) around D=0.0027.
The mechanism of the transition from stable dynamics to unstable dynamics 
arises as noise-induced chaos, which is an universal phenomenon in random dynamical systems (Fig. 5).
The transition point from SS to WS in a parameter space can change depending on noise sequences, so the accurate transition point cannot be detected numerically.

\begin{figure}[ht]    
 \begin{tabular}{cc}
\includegraphics[scale=.20]{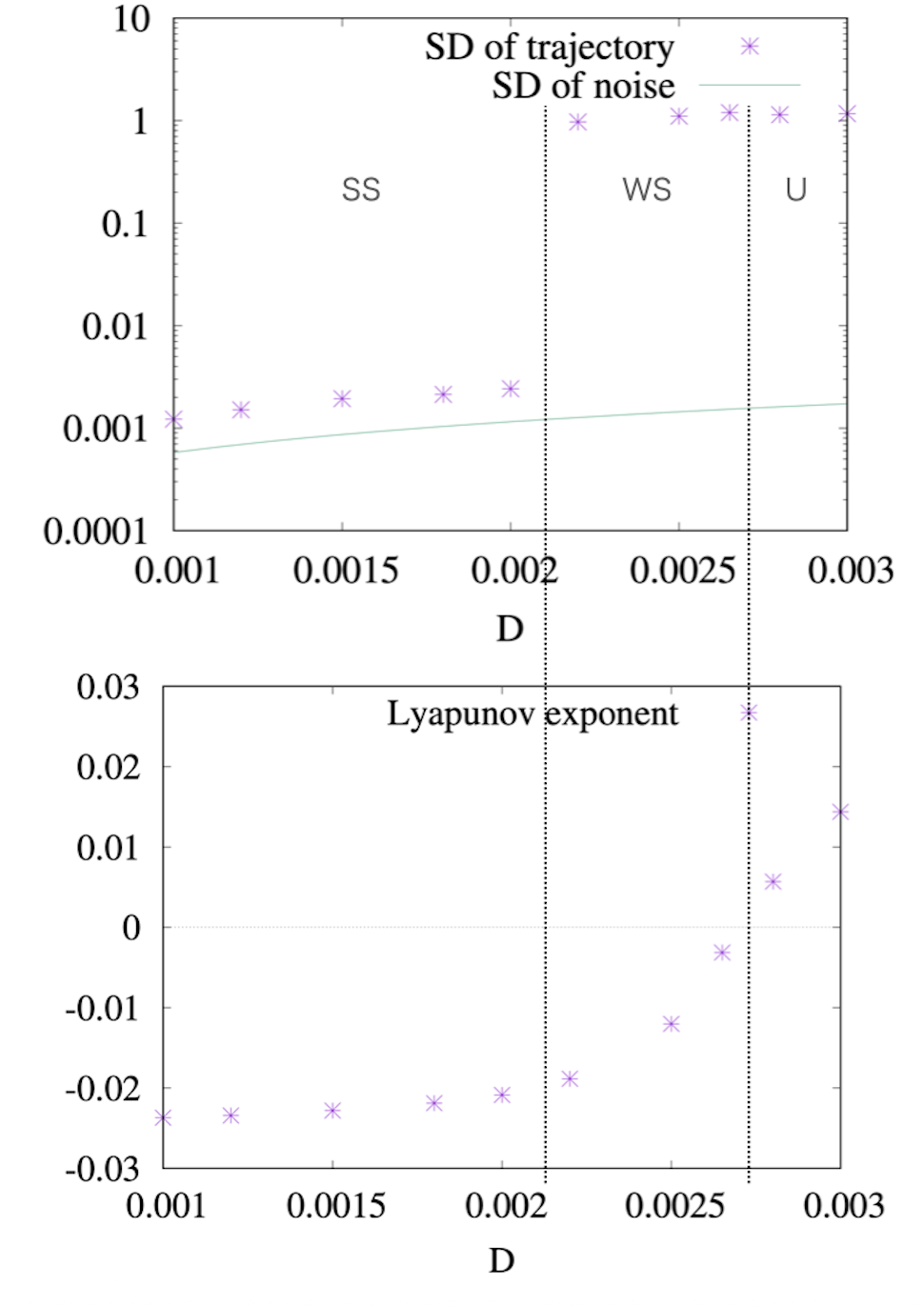}  
\end{tabular} 
\caption{\label{kobayashi_fig:fig1}
   Standard deviation (SD) of a trajectory (top) and the largest Lyapunov exponent (bottom) in the controlled system in changing the system noise intensity $D$. 
   Standard deviation of the trajectory becomes much bigger than that of the system noise above $D=0.0023$. 
   The largest Lyapunov exponent becomes positive above $D=0.0028$.
   The system is strongly stable (SS) in $D<0.0023$,    is weakly stable  (WS), and is unstable (U).
   }       
\end{figure}

\begin{figure}[ht]
    \begin{tabular}{cc}
    \includegraphics[scale=.50]{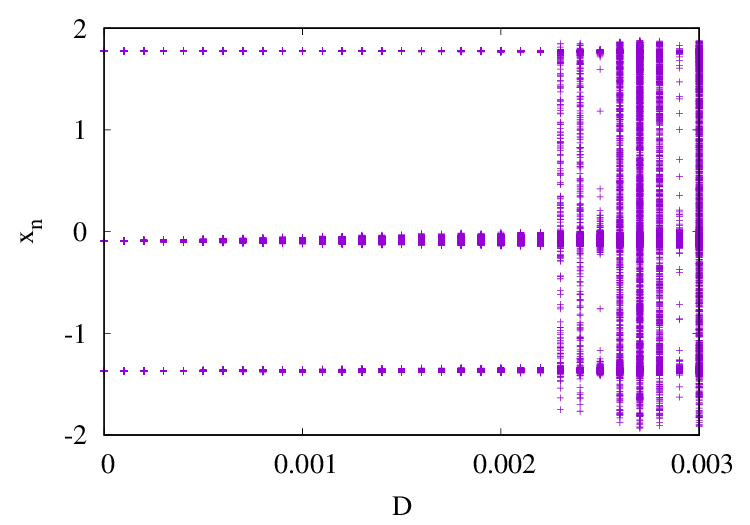} 
\end{tabular} \caption{\label{kobayashi_fig:fig1}
   Bifurcation diagram in changing noise intensity $D$. The transition from SS to WS occurs at $D=0.0023$, which is consistent with Figure 4 (top).
   }       
\end{figure}

\section{Time-delayed feedback control for Random Rossler equation}
In this section, the random R\"ossler equation with time-delayed feedback control is considered:
\begin{eqnarray}
\dot{x}=-y-z+D\xi_x+K(x(t-\tau)-x(t)),\\
\dot{y}=x+ay+D\xi_y+K(y(t-\tau)-y(t)),\\
\dot{z}=b+z(x-c)+D\xi_z+K(z(t-\tau)-z(t)).
\end{eqnarray}

We set the parameters as $a=0.2, b=0.36, c=5.7$. 
This setup is similar to the logistic map in the previous chapter, with parameters that cause chaos to occur near the period-3 window.
The feedback gain is fixed $K=10$.
The delayed time $\tau$ is chosen to match the period of an unstable periodic orbit.  
In this paper, we focus on stabilizing the period-3 unstable periodic orbit with period $17.799$. The stochastic term 
$\xi$ is given by the Gaussian noise (N(0,1)) with the noise intensity $D$.
Note that the standard deviation of system noise $D\xi$ is $D$. 
We calculate numerically Eq.  (5)$\sim$(7) by using Euler-Maruyama method and $\xi$ by the Box-Muller's method with a time step $dt=0.0001$.



We control the system eq. (5) $\sim$ (7) with varying noise intensity $D$.
Fig. 6 shows stabilized periodic orbit (top) and stationary  distribution function for time series projected
onto the $x-y$ plane
(
bottom) for the case $D = 0.1$. 
Since the second Lyapunov exponent is $-0.122$ (the largest exponent is nearly 0), and the standard deviation of trajectory is $0.083$, 
which is the same order as that of the system noise, 
the system is strongly stable (SS).
As D increases, the standard deviation increases proportionally and when $D$ becomes $0.2$, the standard deviation becomes $1.235$, however the second Lyapunov exponent remains at $-0.090$, resulting in the system becoming weakly stable (WS).
Furthermore, when D increases to $0.4$, the standard deviation becomes $1.4234$ and the largest Lyapunov exponent becomes positive at $0.047$, so the system becomes unstable (U).
Fig. 9 shows the transition from strongly stable (SS) to weakly stable (WS) and then to unstable (U).

Fig. 10 shows a bifurcation diagram by changing the noise intensity $D$.
The Poincaré section for the bifurcation diagram is defined by the plane
$
x = 0,\quad \dot{y} < 0.
$
When two successive points in the time series satisfied these conditions,
the intersection is approximated by linear interpolation. 
As the same with the logistic map, which is one of the time-discrete systems, 
loss of stability is caused by noise-induced chaos.

\begin{figure}[ht]
\begin{tabular}{cc}
  \includegraphics[scale=.50]{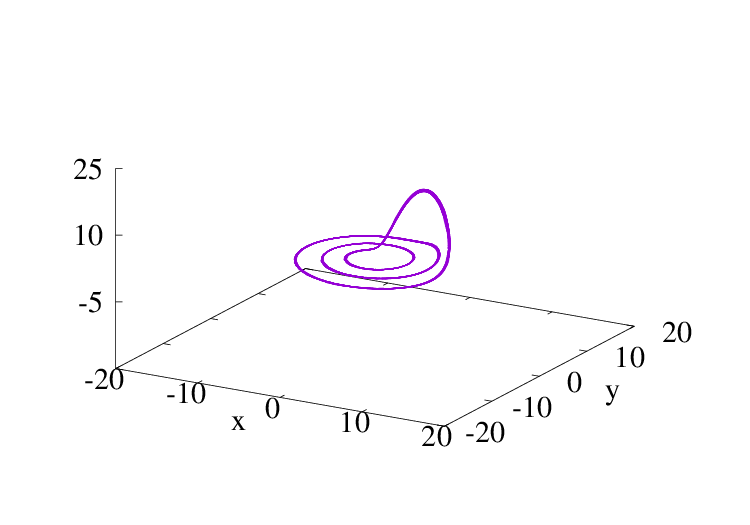} \\
  \includegraphics[scale=.30]{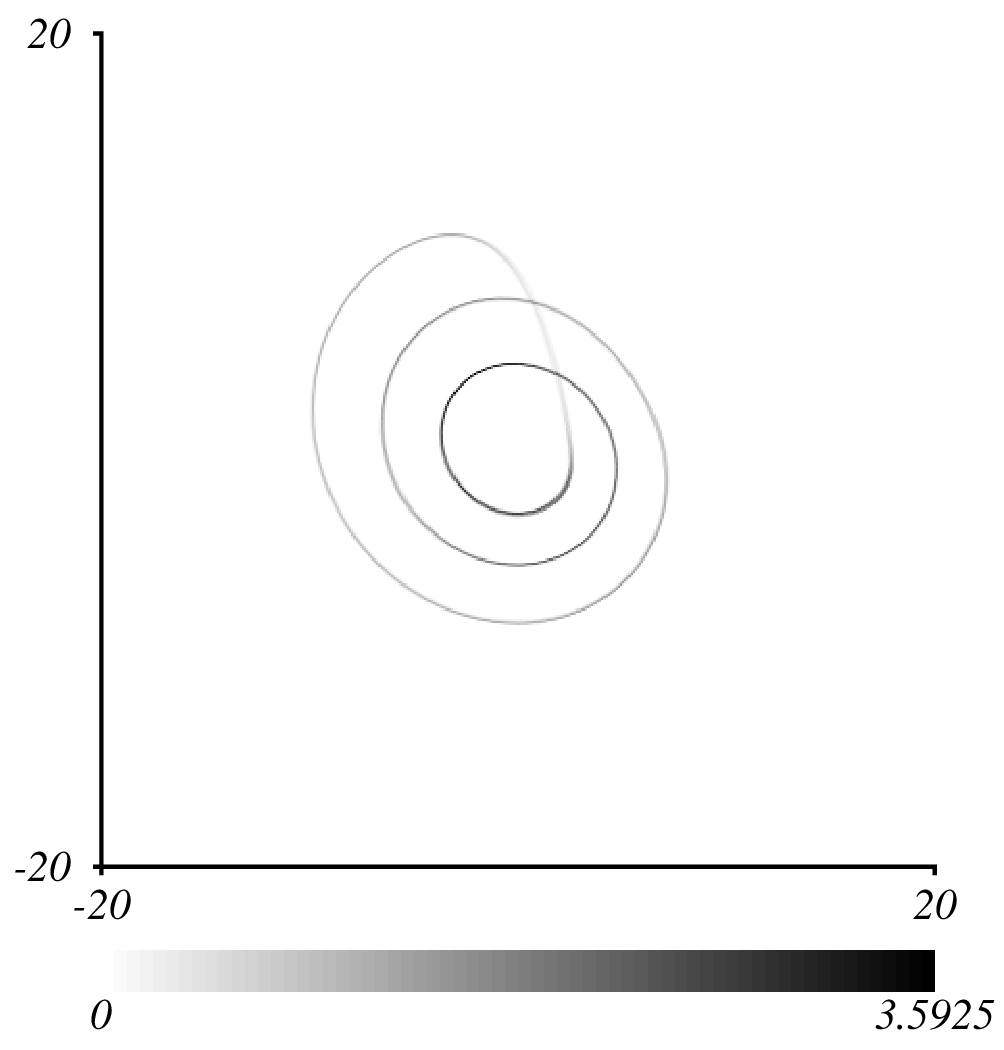} 
      \end{tabular}

    \caption{\label{kobayashi_fig:fig1}
    (left) Stabilized pseudo-periodic orbit with a delayed time $\tau=17.799$ and noise intensity $D=0.1$ and (right) density distribution function for time series projected onto the x-y plane.
    The stabilized orbit becomes nearly indistinguishable from a periodic orbit.
    Standard deviation of trajectory is $0.083$, which is approximately the same order of the standard deviation of system noise $D$. The Largest Lyapunov exponent is trivial ($\lambda_1 \sim 0$). The sencond Lyapunov exponent $\lambda_2$ is $-0.122$.
    This case corresponds to the strongly stable (SS).   
    }       
\end{figure}

\begin{figure}[ht]
\begin{tabular}{cc}
    \includegraphics[scale=.50]{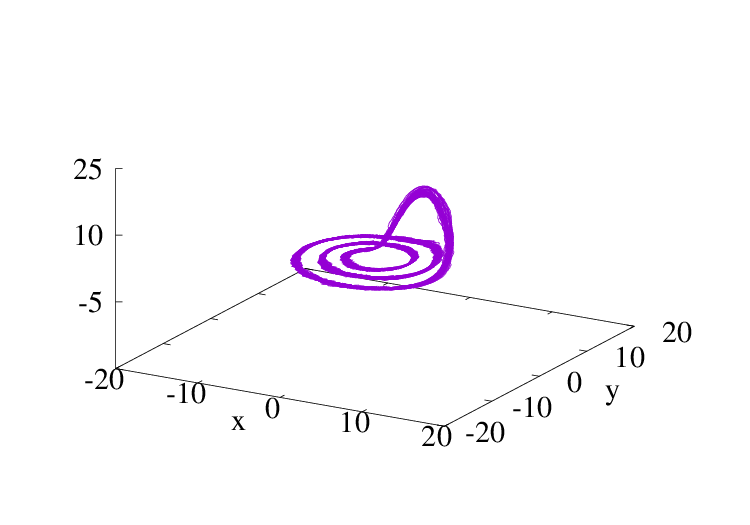} \\
    \includegraphics[scale=.30]{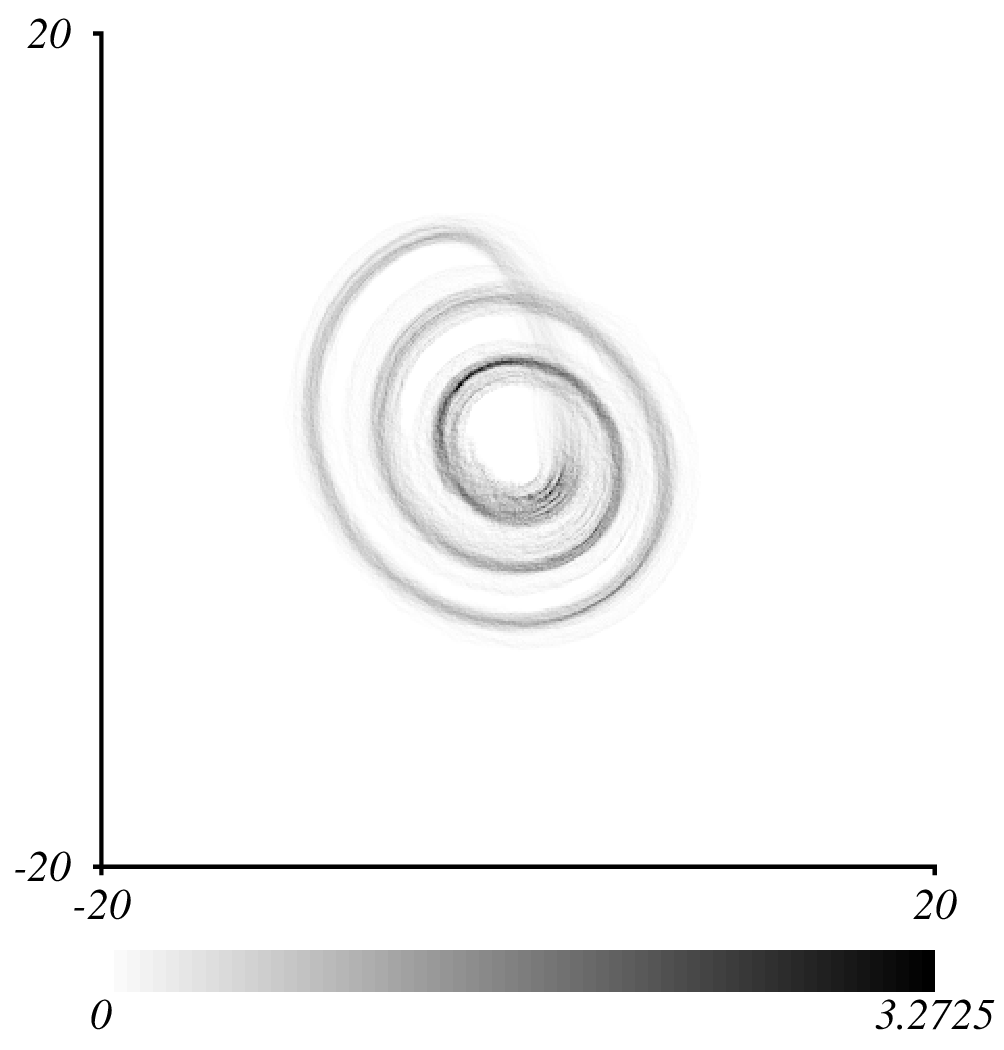} 

      \end{tabular}

    \caption{
    \label{kobayashi_fig:fig1}
    (left) partially chaotic orbit with a delayed time $\tau=17.799$ and noise intensity $D=0.2$ and (right) stationary  distribution function for time series projected onto the x-y plane.
    The stabilized orbit becomes nearly indistinguishable from a periodic orbit.
    Standard deviation of trajectory is $1.235$, which is larger than one of noise. 
    The Largest Lyapunov exponent is trivial ($\lambda_1 \sim 0$). The sencond Lyapunov exponent $\lambda_2$  is $-0.090$.
    This case corresponds to the weakly stable (WS) 
    with $p=0.73$. The value $p$ represents the proportion of time during which $\sqrt{(x_p(t)-x_p(t-\tau))^2+(y_p(t)-y_p(t-\tau))^2}$ does not exceed the standard deviation $D$ of the system noise, where $x_p$ and $y_p$ represent coordinates on the Poincaré section which is defined by the plane $x=0, {\dot y}<0$. 
     }     
\end{figure}

\begin{figure}[ht]
    \begin{tabular}{ccc}
    \includegraphics[scale=.50]{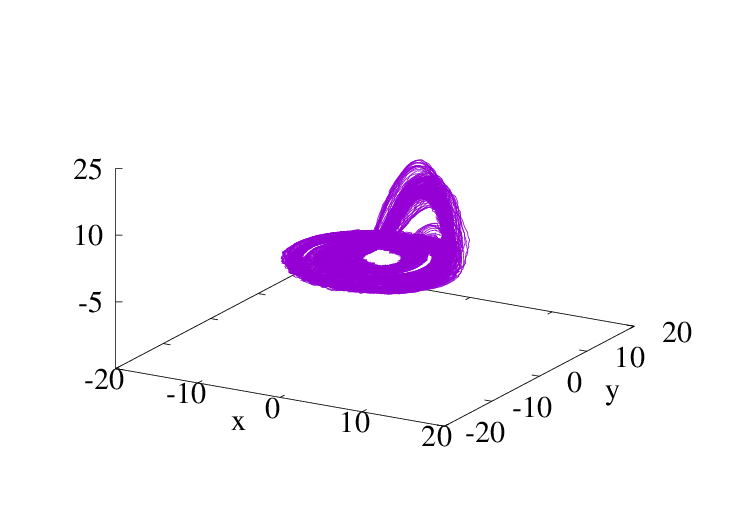} \\
    \includegraphics[scale=.30]{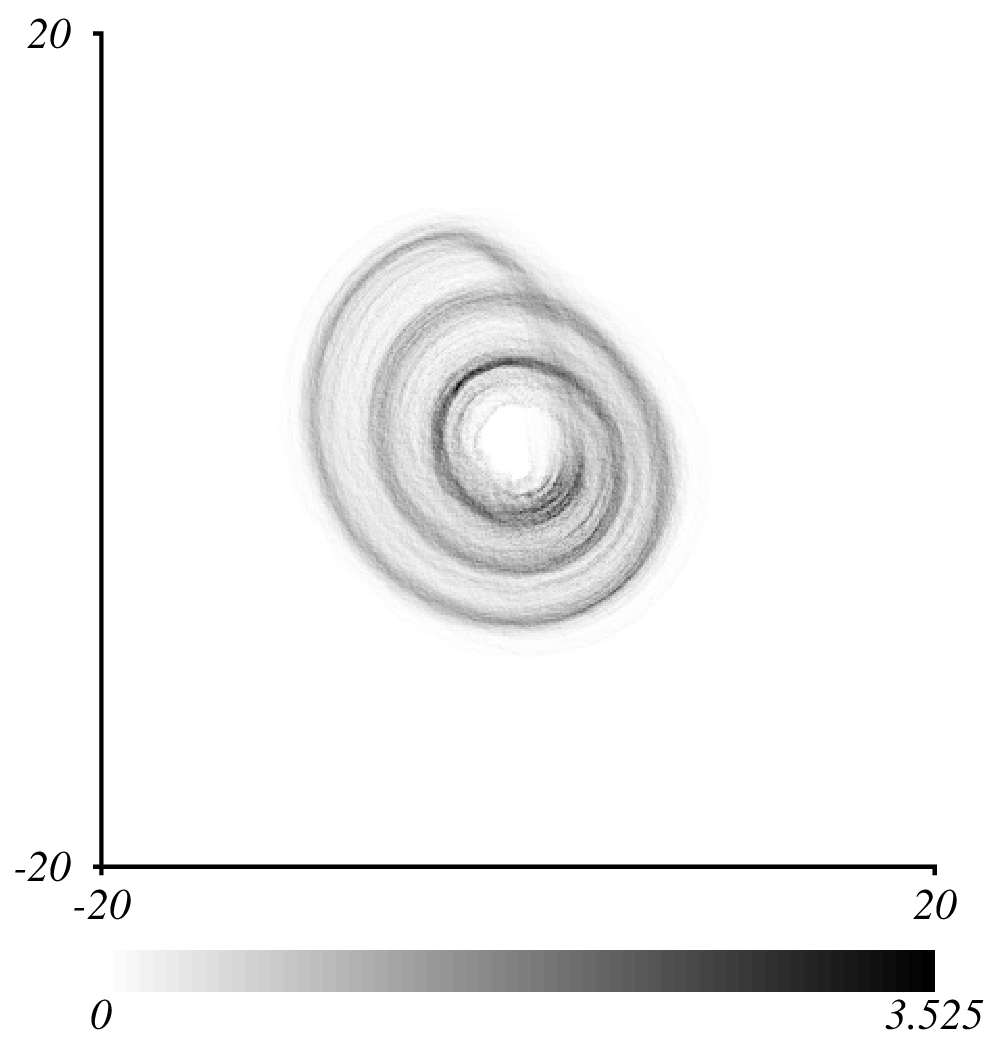} 
      \end{tabular}

    \caption{\label{kobayashi_fig:fig1}
      (left) Random chaotic attractor with the delayed time $\tau=17.799$ and noise intensity $D=0.4$ and (right) density distribution function for time series projected onto the x-y plane.
    Standard deviation of trajectory is $1.435$, which is much larger than one of noise. 
    The Lyapunov exponent is $0.047$.
    This case corresponds to the unstable (U).   
    } 
\end{figure}

   

\begin{figure}[ht]
    \begin{tabular}{cc}
\includegraphics[scale=.20]{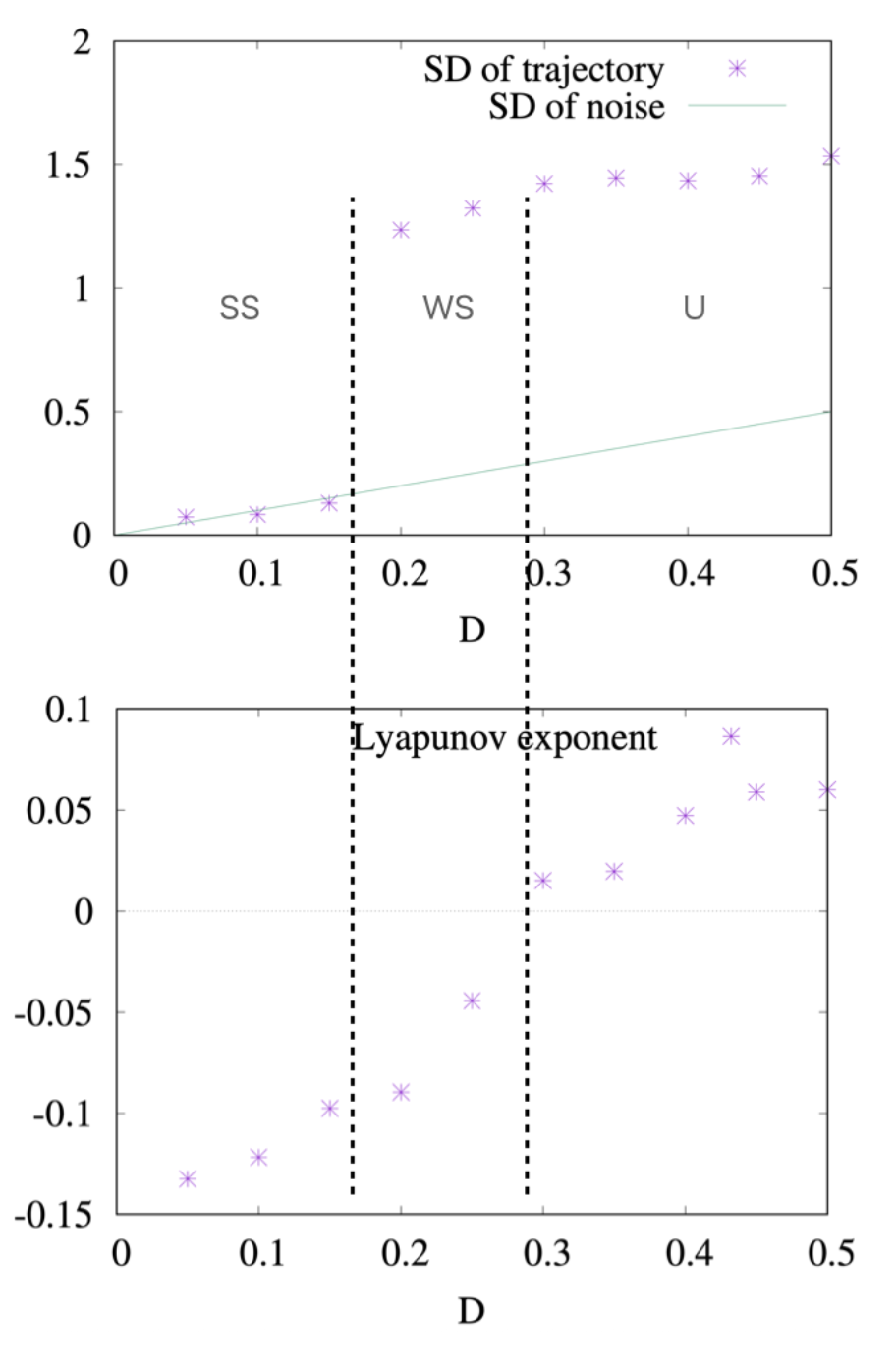} 
\end{tabular}
\caption{\label{kobayashi_fig:fig1}
    Standard deviation of a trajectory (top) and the second Lyapunov exponent (bottom) in the controlled system in changing the system noise intensity $D$ for the Rossler system. 
   Standard deviation of the trajectory becomes much bigger than that of the system noise above $D=0.18$. 
   The largest Lyapunov exponent except the 0-Lyapunov exponent becomes positive above $D=0.3$.
   The system is strongly stable (SS) in $D<0.18$,    is weakly stable  (WS) in $0.18  \le D<0.3$, and is unstable (U) in $D \geq 0.3$.
   }       
\end{figure}


\begin{figure}[ht]
    \begin{tabular}{cc}
  \includegraphics[scale=.50]{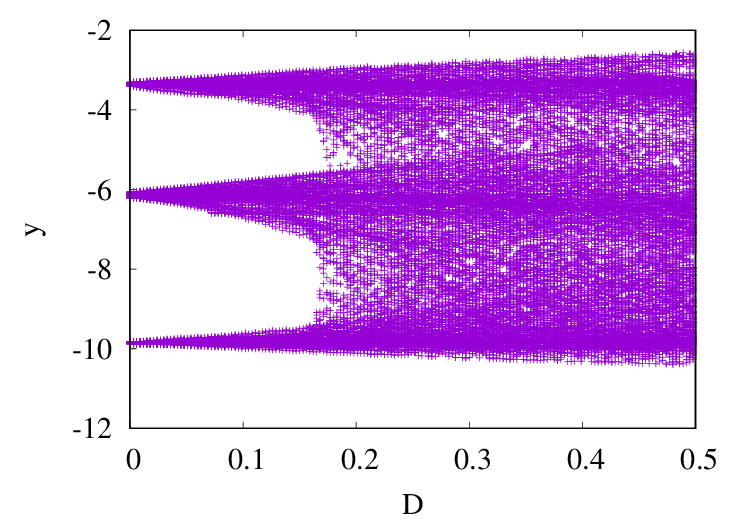} 
\end{tabular} \caption{
   Bifurcation diagram on the Poincare plane, which is defined $x=0, {\dot y}<0$, in changing noise intensity $D$. The control fails at $D=0.3$.
   }       
\end{figure}

\section{Concluding remarks}\label{sec:conclusion}
In this paper, time-delayed feedback control is demonstrated for random dynamical systems, such as the random logistic map and the stochastic  R\"ossler system.

In the time-delayed feedback control of deterministic systems, the only measure of control success is a negative Lyapunov exponent. In controlling chaos for  random dynamical systems, the Lyapunov exponent alone is not sufficient, and orbital fluctuations must be taken into account in order to define successful control. 
We have developed the following three region (see TABLE I): (i) strongly stable  (SS) if the Lyapunov exponent is negative,  and the orbit fluctuation is comparable to that of the system noise, (ii) weakly stable (WS) if the Lyapunov exponent is negative, but the orbit fluctuations are larger than the noise, (iii) unstable  (U) if the Lyapunov exponent is positive.
These definitions consist only of measurable quantities. 
The validity of the definitions was then demonstrated in actual systems, a  random logistic map and a stochastic  R\"ossler system.

In the framework of random dynamical systems, the controlled orbit under TDFC corresponds to a realization of the random pullback attractor. 
In the SS regime, the random attractor degenerates to a small tubular neighborhood around the deterministic periodic orbit. 
In WS, the attractor remains a random perturbation of the periodic orbit but becomes thickened and partially chaotic.  
In contrast, under U the pullback attractor becomes a random strange attractor with a positive Lyapunov exponent.
The SS–WS–U  classification by using Lyapunov exponent and noise variances can be interpreted as a stochastic generalization of the deterministic criterion for the time-delayed feedback control based solely on Lyapunov exponents. 
It is expected to be a general principle for feedback control in noised nonlinear systems, regardless of specific models.

\begin{table*}[htbp]
    \centering
    \caption{Classification of successful control in random dynamical systems with time-delayed feedback. 
    Here $\lambda$ denotes the (almost-sure) Lyapunov exponent and 
    $\sigma$ and $\varepsilon$ denote the standard deviation of the controlled trajectory 
    and the system noise, respectively.}
    \begin{tabular}{c|c|c|c}
        \hline
        Regime & Lyapunov exponent $\lambda$ & Variance ratio $\sigma/\varepsilon$ & Dynamics \\ \hline
        Strongly stable (SS) 
            & $\lambda < 0$ 
            & $\sigma/\varepsilon =O(1)$ 
            & Pseudo-periodic orbit with noise-level fluctuations \\ \hline
        Weakly stable (WS) 
            & $\lambda < 0$ 
            & $\sigma/\varepsilon \gg 1$
            & Thickened pseudo-periodic orbit / partially chaotic orbit \\ \hline
        Unstable (U) 
            & $\lambda > 0$
            & $\sigma/\varepsilon \gg 1$
            & Random strange attractor \\ \hline
    \end{tabular}
    \label{table:SCWCUC}
\end{table*}

In deterministic systems, the loss of stability under TDFC is typically associated with a bifurcation of the targeted periodic orbit.
In contrast, in random dynamical systems the transition from stable (SS/WS) to unstable (U) occurs while the deterministic skeleton of the system still possesses a stable periodic orbit.
This transition is caused by noise-induced chaos: when the noise intensity exceeds a critical level,
the stochastic fluctuations overwhelm the deterministically stable periodic orbit
and induce a random strange attractor with a positive Lyapunov exponent.
In this regime, the feedback signal no longer compensates the phase deviations of the target orbit,
and consequently the control fails even though the feedback remains bounded.

In this paper, 
$p$-stability is introduced as a measure of weak stability in random dynamical systems, based on the sojourn-time rate within the non-chaotic region of the dynamics. 
Whether a weakly stable situation constitutes a control success or failure depends on the context and on the specific problems at hand. The details of this issue will be discussed elsewhere.

\section{Appendix}
\subsection{Random pullback attractor and Lyapunov exponent}

To parametrize a noise realization $\omega \in \Omega$ in time $t$, we introduce a family of measure-preserving maps 
\[
\theta_t : \Omega \to \Omega
\]
satisfying $\theta_0 = \mathrm{id}_\Omega$ and $\theta_{s+t} = \theta_s \circ \theta_t$ for each time $t$.  
The time evolution of random dynamical systems can be described by the stochastic flow
\[
\Phi(t,\omega) : X \to X
\]
satisfying the cocycle property
\[
\Phi(t+s,\omega) = \Phi(t,\theta_s(\omega)) \circ \Phi(s,\omega).
\]

In the case of single attractor systems, a set $A(\omega)$ is called a \emph{random pullback attractor} if it satisfies the following three conditions (see \cite{Arnold1998, Chekroun1685} for the exact definition):

\begin{enumerate}
\item \textbf{Compactness}:  
$A(\omega)$ is compact, i.e.,
\[
A(\omega) := \{ x \in X \mid (\omega, x) \in A(\omega) \} \subset X
\]
is compact for almost all $\omega \in \Omega$.

\item \textbf{$\Phi$-invariance}:  
For all $t$,
\[
\Phi(t,\omega) A(\omega) = A(\theta_t \omega)
\]
for almost all $\omega \in \Omega$.

\item \textbf{Pullback attraction}:  
\[
\lim_{t \to \infty} d_X \big( \Phi(t, \theta_{-t}\omega) B, \; A(\omega) \big) = 0
\]
holds for all $B \subset X$ and for almost all $\omega \in \Omega$,
where $d_X$ is the Hausdorff semi-distance.
\end{enumerate}

A realization $A(\omega)$ of a random pullback attractor is approximated by a snapshot of trajectories evolving from a set of initial values $B$, with a fixed noise realization $\omega$, and with an integration period given by the pullback time $t_p$.  
For a precise numerical computation, a large number of initial points for $B$ and a large pullback time $t_p$ are adopted.

The Lyapunov exponent of stochastic dynamics on random pullback attractors is given by the average expansion rate of perturbations similarly to those of deterministic dynamics.  
For example, the Lyapunov exponent of a one-dimensional stochastic differential equation
\[
dx = f(x) \, dt + \sigma \, dW_t,
\]
where $W_t$ is the Wiener process, 
is given by
$$
\lambda (\omega , x_0) = \lim_{T \rightarrow \infty} \int_0^T f'(\Phi(t, \omega)x_0)dt,
$$
where $\Phi$ represents the stochastic flow,
and $x_0 \in X$ is the initial condition.
Note that the Lyapunov exponent is a random variable in general.
When the whole system is ergodic and has a single attractor,
$\lambda(\omega, x_0)$ is a constant for almost 
all $\omega$ an for all $x_0$.
Therefore, the system possesses a unique Lyapunov exponent, and throughout this paper we simply refer to this quantity as the Lyapunov exponent $\lambda$.

\section{Acknowledgements} 

Y.S. is supported by JSPS Grant-in-Aid for Scientific Research (B), JP No. 21H01002 and  JST Moonshot R\&D Program, No. 
JPMJMS2282-17.

\section*{Data Availability Statement}

The data and code that support the findings of this study are available from the corresponding author upon reasonable request.

\section*{Conflict of interest}
The authors have no conflicts to disclose.


\bibliographystyle{unsrt}

\bibliographystyle{plain}



\begin{thebibliography}{99.}

\bibitem{Romeiras1992} Romeiras, F.J.,  Grebogi, C.,  Ott, E. and Dayawansa, W.P.: Controlling chaotic dynamical \
systems, Physica D, {\bf 58}, 165--192 (1992)

\bibitem{Shinbrot1993} Shinbrot, T., Grebogi, C.,  Ott, E., and Yorke, J.A.: Using small perturbation to control \
chaos, Nature, {\bf 363}, 411--417 (1993)

\bibitem{Pyragas1992} Pyragas, K.: Continuous control of chaos by self-controlling feedback,
Phys. Lett. A, {\bf 170}, 421--428 (1992)

\bibitem{Bielawski1994} Bielawski, S., Derozier, D., and Glorieux, P.:
Controlling unstable periodic orbits by a delayed continuous feedback,
Phys. Rev. E, {\bf 49}, R971 (1994)


\bibitem{Just1997} Just, W., Bernard, T., Ostheimer, M.,Reibold, E., and Benner H.:
Mechanism of time delayed feedback control,
Phys. Rev. Lett., {\bf 78}, 203--206 (1997)

\bibitem{Just1998} Just, W.: Delayed feedback control of periodic orbits in autonomous systems,
Phys. Rev. Lett., {\bf 81}, 562--565 (1998)

\bibitem{Postlethwaite2009}
Postlethwaite C.M.:
Stabilization of Long-Period Periodic Orbits Using Time-Delayed Feedback Control, Siam Journal on Applied dynamical systems, {\bf 8}, 21--39 (2009)

\bibitem{Kobayashi2012} Kobayashi, U.M. and Aihara K.: Delayed feedback control method for dynamical systems with chaotic saddles, AIP Conf. Proc., {\bf 1468}, 207--2015 (2012)

\bibitem{Yamasue2006}
Yamasue, K. and Hikihara, T.:
Control of microcantilevers in dynamic force microscopy using time delayed feedback, Review of scientific instruments {\bf 77} 053703 (2006)

\bibitem{Nakajima1997} Nakajima, H. :
On analytical properties of delayed feedback control of chaos,
Phys. Lett. A, {\bf 232}, 207--210 (1997)

\bibitem{FiedlerPRL} Fiedler, B., Flunkert, V., Georgi, M., H\"{o}vel, P., and Sch\"{o}ll, E.,
Refuting the odd-number limitation of time-delayed feedback control,
Phys. Rev. Lett. {\bf 98}, 114101-1--114101-4 (2007)

\bibitem{Arnold1998}
Arnold, L., {\it Random Dynamical Systems} (Springer Berlin, Heidelberg, 1998).

\bibitem{Chekroun1685}
Chekroun,M.D., Simonnet, E.,and Ghil, M., Physica D: Nonlinear Phenomena {\bf 240}, 1685 (2011).

\bibitem{Janson2004}
Janson, N.B., Balanov, A.G., Sch\"oll, E.,
Delayed feedback as a means of control of noise-induced motion,
Phys. Rev. Lett. {\bf 93}, 010601 (2004)

\bibitem{Balanov2004}
Balanov, A.G., Janson, N.B., Sch\"oll, E.,
Control of noise-induced oscillations by delayed feedback,
Physica D: Nonlinear Phenomena. {\bf 199}, 1 (2004)

\bibitem{Ohira1999}
Ohira, T., and Sato, Y., 
Resonance with noise and delay,
Phys. Rev. Lett. {\bf 82}, 2811 (1999).

\bibitem{Thai2018}
Doan, T.S., Engel, M., Lamb, J.S.W., and Rasmussen, M., Hopf bifurcation with additive noise,Nonlinearity {\bf 31}, 4567 (2018).

\bibitem{Sato2018}
Sato, Y.,Doan, T.S., Engel, M., Lamb, J.S.W., and Rasmussen, M., 	
Dynamical characterization of stochastic bifurcations in a random logistic map,
arXiv:1811.03994.

\end{thebibliography}

\end{document}